\documentclass[reprint,amsmath,amssymb,aps]{revtex4-1}
\usepackage{graphicx}
\usepackage{amsmath}
\usepackage{epsfig}
\usepackage[colorlinks]{hyperref}
\usepackage{siunitx} 
\usepackage{ulem} 

\begin{document}

\title{Quantum process tomography of a controlled-phase gate for time-bin qubits}
\author{Hsin-Pin Lo}
\email{hsinpin.lo.cn@hco.ntt.co.jp}
\affiliation{NTT Basic Research Laboratories, NTT Corporation, 3-1 Morinosato Wakamiya, Atsugi, Kanagawa, 243-0198, Japan}
\author{Takuya Ikuta}
\affiliation{NTT Basic Research Laboratories, NTT Corporation, 3-1 Morinosato Wakamiya, Atsugi, Kanagawa, 243-0198, Japan}
\author{Nobuyuki Matsuda} \email{Present address: Department of Communications Engineering, Graduate School of Engineering, Tohoku University}
\affiliation{NTT Basic Research Laboratories, NTT Corporation, 3-1 Morinosato Wakamiya, Atsugi, Kanagawa, 243-0198, Japan}
\author{Toshimori Honjo}
\affiliation{NTT Basic Research Laboratories, NTT Corporation, 3-1 Morinosato Wakamiya, Atsugi, Kanagawa, 243-0198, Japan}
\author{William J. Munro}
\affiliation{NTT Basic Research Laboratories, NTT Corporation, 3-1 Morinosato Wakamiya, Atsugi, Kanagawa, 243-0198, Japan}
\author{Hiroki Takesue}
\email{hiroki.takesue.km@hco.ntt.co.jp}
\affiliation{NTT Basic Research Laboratories, NTT Corporation, 3-1 Morinosato Wakamiya, Atsugi, Kanagawa, 243-0198, Japan}
\date{\today}						

\begin{abstract}
Time-bin qubits, where information is encoded in a single photon at different times, have been widely used in optical fiber and waveguide based quantum communications. With the recent developments in distributed quantum computation, it is logical to ask whether time-bin encoded qubits may be useful in that context. We have recently realized a time-bin qubit controlled-phase (C-Phase) gate using a 2$\times$2 optical switch based on a lithium niobate waveguide, with which we demonstrated the generation of an entangled state. However, the experiment was performed with only a pair of input states, and thus the functionality of the C-Phase gate was not fully verified. In this research, we used quantum process tomography to establish a process fidelity of 97.1$\%$. Furthermore, we demonstrated the controlled-NOT gate operation with a process fidelity greater than 94$\%$. This study confirms that typical two-qubit logic gates used in quantum computational circuits can be implemented with time-bin qubits, and thus it is a significant step forward for realization of distributed quantum computation based on time-bin qubits.
\end{abstract}

\maketitle
\section{introduction}
We have entered an era where quantum physics allows us to perform information processing tasks that would be impossible in a world governed only by classical physics \cite{Nielsen_Chuang,quantumcomputer, chemistry}. For instance, the quantum laws of nature offer the possibility of secure communications, quantum sensing, imaging and metrology beyond the classical limit, and computing capabilities exponentially faster than any conventional digital machine. In particular, central to the operation of quantum computers are quantum bits (qubits) interacting with each other via quantum gates. A wide range of physical systems, including ion traps \cite{steane_iontrap}, superconducting circuits \cite{QST_superconducting_IBM}, quantum dots \cite{Li_Qdot_gate, Hanson_dot_RMP_2007, Zwanenburg_dot_RMP_2007}, and photonic qubits based on polarization \cite{polspdc, beamlikeLo, beamlike2x2}, temporal modes \cite{qst_timebin}, and frequency modes \cite{frequencybinqubit} have shown themselves to be excellent qubits that can be easily manipulated and have long coherence times. In a number of these systems, quantum logic gates \cite{ion, gate_ion_trap, phasequbit, gate_superconducting, superconducting, cnot_gate_superconducting, solid, frequencybinCNOT} have been used to manipulate qubits and run small quantum algorithms. Here, while photons would seem an unusual choice for quantum computation due to the difficulty of their interacting with each other via nonlinear interactions, schemes for linear optical quantum computation using only linear optics elements and postselection have been proposed \cite{KLM,postselection}. In fact,  two-qubit entangling gates including controlled-phase (C-Phase) \cite{cphasetheory} and controlled-NOT (CNOT) \cite{cnottheory} gates have been demonstrated using path \cite{gate_photon_OBrien} and polarization \cite{cphase_pdbs, gate_Bell, kok_RMP} encoded photonic qubits. 

Photons have been used for decades as information carriers in optical fibers and waveguides. The quantum internet \cite{qinternet} requires multiple nodes, each with the ability to perform small-scale quantum processing. Over long distances, the primary method of operating quantum networks is to use optical networks and photon-based qubits. Photonic quantum logic gates are useful for even more advanced quantum communications, such as a quantum repeater system without using quantum memories based on quantum error correction \cite{Bill_repeater}. Such networks will rely on optical fiber links, making information processing. However, the polarization degrees of freedom, for instance, are difficult to preserve through such quantum channels because they are strongly affected by fluctuations in the birefringence and refractive indices of the fibers. Photons, however, possess many other degrees of freedom that could be enlisted for this task.

In such a role, however, it is important to choose the right degrees of freedom to connect to each other. Time-bin qubits at telecommunications wavelengths, on the other hand, are especially promising as quantum information carriers through optical fiber. They have been used for many quantum communications tasks, including quantum key distribution over extremely long distances \cite{time-energy_Brendel_1999, time-energy_Gisin_2007, Honjo_timebin_QKD}.  For distributed quantum computation, two-qubit logic gates for time-bin qubits are needed to entangle independent states through long optical fibers. Recently, we demonstrated the generation of an entangled state using a C-Phase gate for time-bin qubits, which is a high-speed two-input, two-output (2$\times$2) optical switch based on a lithium niobate waveguide modulator \cite{cphasetimebin}. 

In this work, we experimentally implemented a two time-bin qubit C-Phase gate and fully characterized its operation through quantum process tomography (QPT). We reconstructed its process matrix, which clearly showed the C-Phase gate operation, and determined its entangling capability. Further we also demonstrated a high fidelity time-bin qubit CNOT gate using complementary computation bases.

\begin{figure*}[t]
\centering 
\includegraphics[width=18cm]{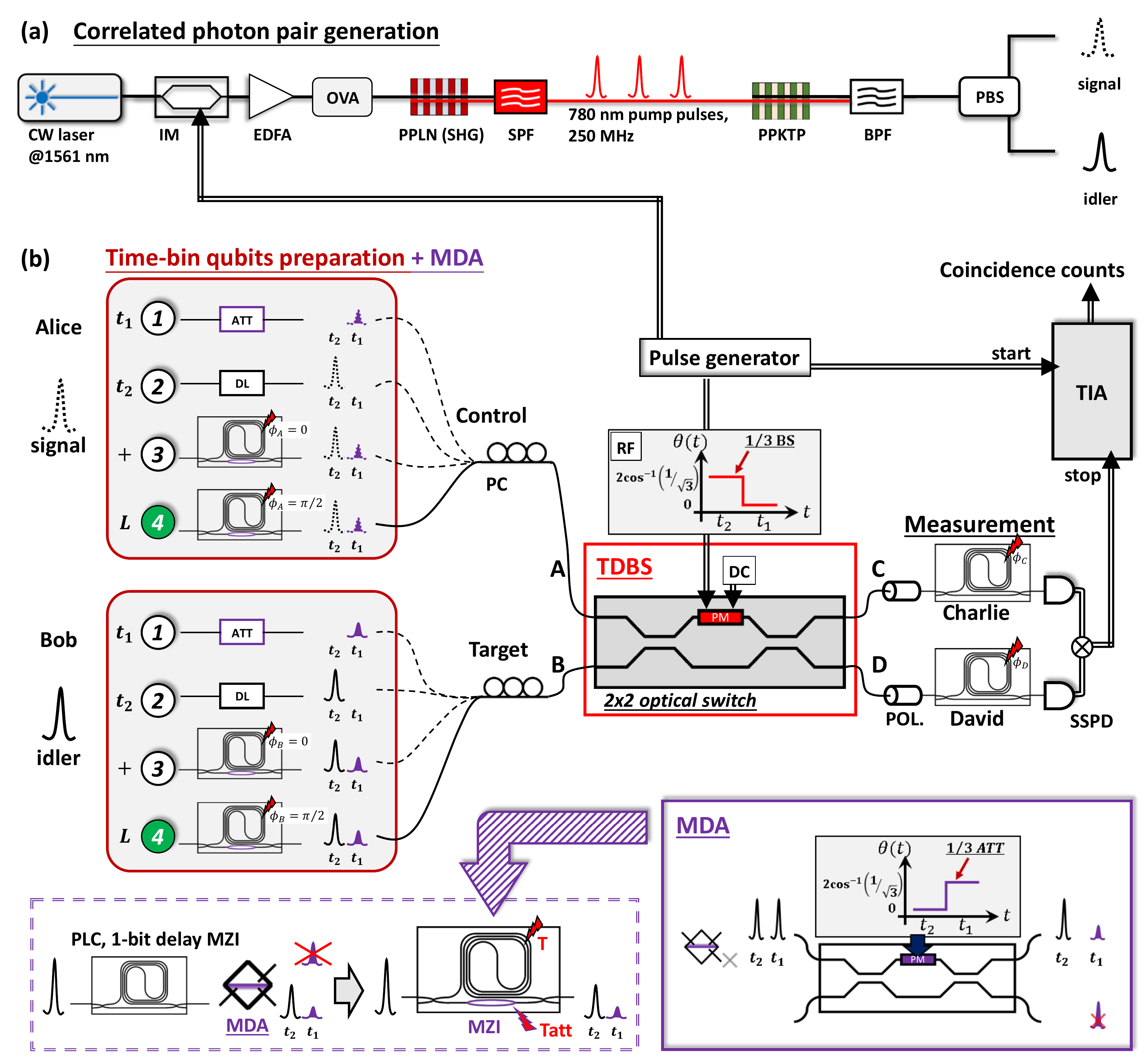}
\caption{Experimental setup for the implementation and quantum process tomography measurement of our time-bin qubit C-Phase gate, with schematic illustrations of the time-dependent beam splitter (TDBS) and mode-dependent attenuator (MDA) operations using a 2$\times$2 optical switch. (a) Correlated photon pair generation. (b) Four methods of time-bin qubits preparation for TDBS and quantum state tomography measurement. CW laser, continuous wave laser; IM, intensity modulator; EDFA, erbium-doped fiber amplifier; OVA, optical variable attenuator; PPLN, periodically poled lithium niobate waveguide; SHG, second harmonic generation; PPKTP, periodically poled potassium titanyl phosphate waveguide; PBS, polarization beam splitter; PLC, planar lightwave circuit; MZI, Mach-Zehnder interferometer; Circles 1, 2, 3, and 4, preparation methods; PC, polarization controller; POL., polarizer; PM, electro-optic phase modulator; T, temperature controller for relative phase ($\phi_y$) adjusting; Tatt, temperature setting of MZI in short arm of PLC by independent temperature controller; SSPD, superconducting single-photon detector; TIA, time interval analyser; ATT, one-third amplitude attenuator; DL, delay line; MDA, mode-dependent attenuator.}
\label{expqpt}
\end{figure*}

\section{Controlled-phase gate for time-bin qubits}
\label{concept}
An ideal C-Phase gate applies a relative $\pi$-phase shift only when both input qubits are $\left|1 \right\rangle$; otherwise, it does nothing \cite{Nielsen_Chuang}. In the case of polarization qubits \cite{cphasetheory}, one realization of the C-Phase gate consists of two processes: a one-third beam splitter only for vertically polarized photons and mode-dependent attenuation to horizontally polarized photons. This scheme was demonstrated using a polarization dependent beam splitter \cite{cphase_pdbs}. Similar to the case of a polarization qubit gate, we design a time-bin qubit C-Phase gate operation with a mode-dependent attenuator (MDA) and time-dependent beam splitter (TDBS) by using a 2$\times$2 optical switch, as shown in Fig.~\ref{expqpt}. The 2$\times$2 optical switch  is a lithium-niobate-waveguide Mach-Zehnder interferometer (MZI) that includes an electro-optic phase modulator (PM) \cite{takesue_switch} in one of its optical paths. 

Here, we introduce the concept of our C-Phase gate for time-bin qubits. First, Alice and Bob prepare time-bin qubits as control and target states for the 2$\times$2 switch, which are represented as $\left|\psi \right\rangle_A = n_{1A} \left|t_1 \right\rangle_A + n_{2A} e^{i \phi_A} \left|t_2 \right\rangle_A$ and $\left|\psi \right\rangle_B = n_{1B} \left|t_1 \right\rangle_B + n_{2B} e^{i \phi_B} \left|t_2 \right\rangle_B$, respectively, where $n_{xy}$ denotes the amplitude of $\left|t_{x}\right\rangle_y$ which is a positive real number, satisfying $n_{1y}^2+n_{2y}^2=1$, where $\phi_y$ is the phase difference between temporal positions $t_1$ and $t_2$. Here, we encode a logical 0 (1) in the time-bin state $\left|t_{1}\right\rangle$ ($\left|t_{2}\right\rangle$), which is a single photon in the first (second) temporal mode. By applying a time-varying signal to the PM of the 2$\times$2 switch, the splitting ratio changes in time. The evolution of a single time-bin qubit with the 2$\times$2 switch is described by \cite{takesue_switch}
\begin{eqnarray}
\begin{split}
\left|t_k \right\rangle_A&=\cos\left(\frac{\theta(t_k)}{2}\right)\left|t_k \right\rangle_C-\sin\left(\frac{\theta(t_k)}{2}\right)\left|t_k \right\rangle_D, \\ 
\left|t_k \right\rangle_B&=\sin\left(\frac{\theta(t_k)}{2}\right)\left|t_k \right\rangle_C+\cos\left(\frac{\theta(t_k)}{2}\right)\left|t_k \right\rangle_D, 
\label{sw1}
\end{split}
\end{eqnarray}
where $\theta(t_k)$ represents the phase difference between two arms of the MZI at time $t_k$ while the indices $C$ and $D$ indicate the two output ports of the interferometer. For the C-Phase gate operation, we set $\theta(t_1)=0$ and $\theta(t_2)=2\cos^{-1}(\frac{1}{\sqrt{3}})$, which means that the 2$\times$2 optical switch passes the first temporal mode and works as a one-third beam splitter for the second mode. 

By performing a coincidence measurement between Charlie and David (see Fig.~\ref{expqpt}(b)), we obtain the state
\begin{eqnarray}
\label{flip1}
&&n_{1A} n_{1B} \left|t_1t_1 \right\rangle_{CD} + n_{1A} n_{2B}e^{i\phi_B}\sqrt{\frac{1}{3}}\left|t_1t_2 \right\rangle_{CD} \\
&+ & n_{2A} n_{1B}e^{i\phi_A} \sqrt{\frac{1}{3}}\left|t_2t_1 \right\rangle_{CD} - n_{2A} n_{2B}e^{i(\phi_A+\phi_B)} \frac{1}{3}\left|t_2t_2 \right\rangle_{CD} \nonumber. 
\end{eqnarray}
Here, only the sign of $\left|t_2t_2 \right\rangle$ is flipped by the C-Phase gate operation. The amplitude unbalance can be eliminated by applying 1/3 attenuations only to the $t_1$ mode before or after the 2$\times$2 switch, as in the case of previous linear optics controlled gates based on polarization and spatial mode qubits \cite{gate_photon_OBrien,cnottheory,cphasetheory, cphase_pdbs}. Such mode-dependent attenuations can be implemented using additional optical switches operated with phase conditions $\theta(t_1)=2\cos^{-1}(\frac{1}{\sqrt{3}})$ and $\theta(t_2)=0$,  respectively, as shown in Fig.~\ref{expqpt}. After this operation, the output state for the C-Phase gate operation is given by
\begin{eqnarray}
&&n_{1A} n_{1B} \left|t_1t_1 \right\rangle_{CD} + n_{1A} n_{2B} e^{i \phi_B} \left|t_1t_2 \right\rangle_{CD} \nonumber \\ 
&+& n_{2A} n_{1B} e^{i \phi_A} \left|t_2t_1 \right\rangle_{CD} -  n_{2A} n_{2B} e^{i(\phi_A+\phi_B)} \left|t_2t_2 \right\rangle_{CD}.
\label{flip2}
\end{eqnarray}
From Eq.~(\ref{flip1}) and~(\ref{flip2}), because of the postselection and pre-amplitude compensation, the maximum theoretical success probability of the ideal C-Phase gate based on this scheme is 1/9 without optical component loss. This is the same success probability seen in the polarization-based case \cite{cphasetheory,cnottheory}.

\begin{figure*}[t]
\centering 
\includegraphics[width=18cm]{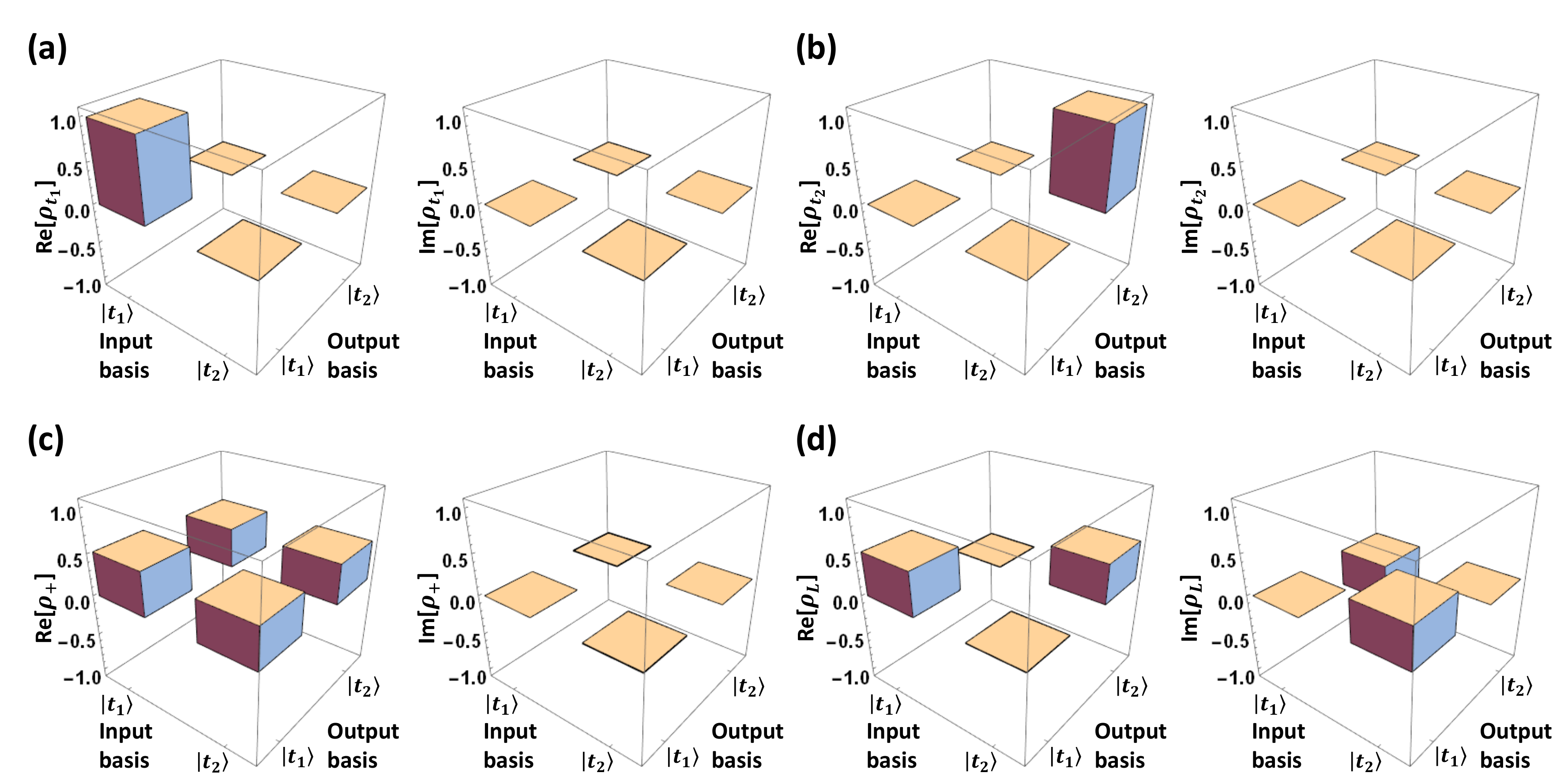}
\caption{Density matrices of single time-bin qubits prepared in the states (a) $\left|t_1 \right\rangle$, (b) $\left|t_2 \right\rangle$, (c) $\left|+ \right\rangle$, and (d) $\left|L \right\rangle$. The left and right parts are the real and imaginary parts of each experimental density matrices.}
\label{singlequbit}
\end{figure*}

\section{Experimental setup}
Our experimental setup is shown in Fig.~\ref{expqpt}, where a 1561-nm continuous-wave (CW) light from an external-cavity diode laser is modulated into a pulse train with a 250-MHz repetition frequency using an optical intensity modulator (IM). We started with a 1.5-$\mu$m light so that we could use external modulation to generate pulses. With external modulation, we can flexibly adjust the repetition frequency of the pump pulses. In the 1.5-$\mu$m telecom band, we can use IMs developed for optical fiber communications, with which high-speed and stable modulation is possible. The pulse generator preparies the synchronized signal to drive the IM and optical switch as well as the start pulse for a time interval analyser (TIA). Thus, we can successfully synchronize the input state, 2$\times2$ optical switch, and measurement.

The pulsed light is amplified by an erbium-doped fiber amplifier (EDFA). The amplified light is then input into a periodically poled lithium niobate (PPLN) waveguide, where a 780.5-nm pulse train is generated via second harmonic generation (SHG). A short pass filter (SPF) is placed after the PPLN waveguide to remove the input 1561-nm pulse light. The quantum correlated photon pairs are generated by pumping the type-II periodically poled potassium titanyl phosphate (PPKTP) waveguide with SHG pulse light, through the process of spontaneous parametric down-conversion (SPDC). The photons are passed through a band pass filter (BPF) with a central wavelength of 1561 nm and a bandwidth of 1.4 nm to reduce noise. By adjusting the optical variable attenuator (OVA) placed after the EDFA, the average number of correlated photon pairs per pulse was set at 0.028 \cite{CAR}. The photon pairs are launched into a polarization beam splitter (PBS), where the pairs are separated into signal and idler photons and sent to Alice and Bob for them to prepare control and target time-bin qubits, respectively.

Alice and Bob launch the control and target qubits into input ports $A$ and $B$ of the 2$\times$2 optical switch (EO Space). By adjusting the DC bias and RF modulation signal to the PM in the switch, the 2$\times$2 switch works as a one-third beam splitter for the $t_2$ mode and passes the $t_1$ mode. The output photons from ports $C$ and $D$ of the switch are sent to the corresponding 1-bit delay interferometers of Charlie and David. The photons output from the interferometers are received by two superconducting single-photon detectors (SSPDs). The detection signals from the SSPDs are used as stop pulses for the TIA where the coincidence signals are counted. The detection efficiencies of the SSPDs of Charlie and David are 57$\%$ and 62$\%$, respectively, and the dark counts for both detectors are less than 40 cps. The coincidence count rate of the QST measurement was $\sim$0.12 Hz. In order to erase the polarization distinguishability of the photon pairs, polarization controllers (PC) are located in front of the input ports of the 2$\times$2 switch, and polarizers (POL.) are placed after the outputs of the optical switch. The insertion loss of the interferometers and optical switch are approximately 2.0 and 7.7 dB, respectively.

\begin{figure*}[t]
\centering 
\includegraphics[width=18cm]{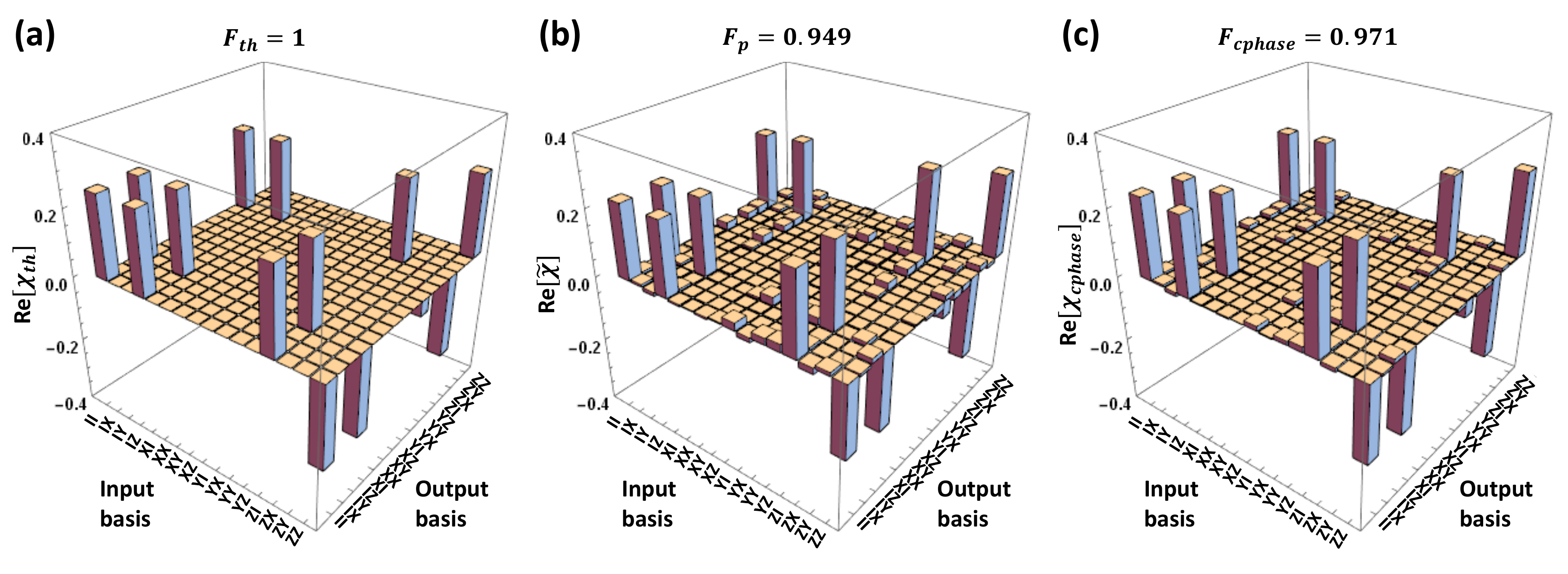}   
\caption{Process matrix of the C-Phase gate for time-bin qubits. (a) Ideal process matrix $\chi_{th}$. (b) Physical process matrix $\tilde{\chi}$ obtained by applying MLE to process matrix $\chi_{exp}$. (c) C-Phase gate process matrix $\chi_{cphase}$ including compensation for the imperfect experimental input time-bin qubits. The imaginary components of the gate are approximately zero and not shown.}
\label{qptmatrixre}
\end{figure*}

\section{Preparation of input states}  
\label{prepare}
To confirm the quality of our input time-bin qubits, we first performed quantum state tomography (QST) \cite{qst, qst_timebin} for each input states without the 2$\times$2 optical switch operation. The experimental density matrices of input state $\left|t_1 \right\rangle$, $\left|t_2 \right\rangle$, $\left|+ \right\rangle$, and $\left|L \right\rangle$, where $\left|+ \right\rangle=\frac{1}{\sqrt{2}} (\left|t_1 \right\rangle + \left|t_2 \right\rangle)$ and $\left|L \right\rangle=\frac{1}{\sqrt{2}} (\left|t_1 \right\rangle +i \left|t_2 \right\rangle)$ \cite{qst_timebin} are shown in Fig.~\ref{singlequbit}. The state fidelities are $F_{t_1}$=99.9 $\pm$ 0.87$\%$, $F_{t_2}$=99.9 $\pm$ 0.87$\%$, $F_{+}$=97.9 $\pm$ 1.44$\%$, and $F_{L}$=98.2 $\pm$ 1.41$\%$ \cite{qst, takesue_teleport_2015}, respectively. The imperfections of superposition states in the fidelity include the imperfect temperature setting of the relative phase of the 1-bit delay interferometers, the dark counts, and multiphoton emission.

As discussed in section~\ref{concept}, additional amplitude attenuation of the $t_1$ mode is necessary to implement the C-Phase gate with MDA. In our setup, the 1/3 amplitude attenuation operation is merged with the state preparation processes of Alice and Bob, which effectively reduces the system loss. Here, we introduce how to prepare the time-bin qubits with methods 1, 2, 3, and 4 shown in Fig.~\ref{expqpt}(b). First, method 1 is used to prepare single temporal state $\left|t_1 \right\rangle$ by inserting a 1/3 attenuator (ATT) to adjust the amplitude of the $t_1$ temporal mode, and method 2 to prepare single temporal state $\left|t_2 \right\rangle$ by adjusting the temporal position of a single photon with an optical delay line (DL). Methods 3 and 4 are used to prepare time-bin qubits $\left|+ \right\rangle$ and $\left|L \right\rangle$ by launching a photon into a 1-bit delay interferometer (fabricated using planar lightwave circuit (PLC) technologies \cite{takesue_plc}), where the relative phase ($\phi_y$) $0$ and $\pi/2$ is adjusted by setting the temperature of the two temporal modes with a temperature controller. Since the phase setting is important for our time-bin qubit demonstration, we choose a more stable temperature controller (Newpoet: LDT-5910C-100V).

As shown in the inset surrounded by the dashed lines in Fig.~\ref{expqpt}, the MDA is equipped with an additional MZI in the short arm of the PLC, and the amplitude attenuation is set to one-third by using an independent temperature controller at the additional MZI. In our QPT experiment, Alice and Bob need to prepare control and target input states by connecting each of the four methods to the 2$\times$2 optical switch one by one, respectively. The prepared time-bin qubits for the input ports of 2$\times$2 switch are represented as $\left|\psi \right\rangle_y \propto \frac{n_{1y}}{\sqrt{3}} \left|t_1 \right\rangle_y + n_{2y} e^{i\phi_y} \left|t_2 \right\rangle_y$.

\section{Characterization of the experimentally realized time-bin qubit C-Phase gate}
The process is expressed by a superoperator $\mathcal{E}$, which represents a quantum gate acting on an arbitrary input state $\rho_{in}$ \cite{Nielsen_Chuang,QPT_white2007}:
\begin{eqnarray}
\mathcal{E}(\rho_{in})=\sum_{m,n}\chi_{mn}\widehat{A}_m\rho_{in} \widehat{A}_n ^\dagger,
\label{qptchi}
\end{eqnarray}
where $\widehat{A}_{m}$ are a basis for operators acting on $\rho_{in}$.
The matrix $\chi$ completely describes the process of the gate operation which is a positive-definite Hermitian matrix by definition. Figure~\ref{qptmatrixre}(a) shows the theoretical process matrix $\chi_{th}$ which represents the ideal C-Phase gate for time-bin qubits of $\widehat{CP}_{ideal}=(I \otimes I+I \otimes Z+Z \otimes I-Z \otimes Z )/2$, where $I$ is the identity matrix while $\{X, Y, Z\}$ are the Pauli matrices. 

For a $d$-dimensional system, we need to prepare $d^2$ pure quantum states as input and to obtain $d^2 \times d^2$ corresponding density matrices by performing QST \cite{qst, qst_timebin}. For our QPT experiments, Alice and Bob choose pure input quantum states $\left|t_1 \right\rangle$, $\left|t_2 \right\rangle$, $\left|+ \right\rangle$, and $\left|L \right\rangle$, which correspond to time-bin qubit preparation methods 1, 2, 3, and 4 shown in Fig.~\ref{expqpt}(b), respectively. Thus, they can prepare 16 distinct input states--$\left|t_1t_1 \right\rangle$, $\left|t_1t_2 \right\rangle$, $\left|t_1+ \right\rangle$, $\left|t_1L \right\rangle$, $\left|t_2t_1 \right\rangle$, $\left|t_2t_2 \right\rangle$, $\left|t_2+ \right\rangle$, $\left|t_2L \right\rangle$, $\left|+t_1 \right\rangle$, $\left|+t_2 \right\rangle$, $\left|++ \right\rangle$, $\left|+L \right\rangle$, $\left|Lt_1 \right\rangle$, $\left|Lt_2 \right\rangle$, $\left|L+ \right\rangle$, and $\left|LL \right\rangle$--for the C-Phase gate operation. We then performed QPT using those input states run through our C-Phase gate to reconstruct process matrix $\chi_{exp}$ defined in Eq.~(\ref{qptchi}). 

As we know, the eigenvalues of the physical process matrix must all be positive. However, the eigenvalues of process matrix reconstructed from the experimental data has several slightly negative values, which implies that the process matrix is unphysical \cite{qptcnot}. To overcome this problem, we used maximum likelihood estimation (MLE) \cite{qptcnot,mle} to obtain the physical process matrix $\tilde{\chi}$ shown in Fig.~\ref{qptmatrixre}(b). The process fidelity $F_p$ in this case was 94.9 $\pm$ 5.6$\%$ \cite{error}. 

Although the physical process matrix $\tilde{\chi}$ consists of an input state and C-Phase gate operation, a photonic polarization qubit experiment can fully characterize the gate operation because of the input state provides high state fidelity \cite{cphase_pdbs, gate_Bell, qptcnot}. Unfortunately, as shown in Sec.~\ref{prepare}, our input time-bin state preparation is imperfect, so we needed to separate the imperfections from our C-Phase gate operation. First, we reconstruct the process matrix $\chi_{input}$ by using the experimental input states and ideal C-Phase gate operation with QPT. The process fidelity $F_{input}$ is 98.5$\%$. Because the C-Phase gate is perfect, $\chi_{input}$ can present the imperfection of the input states. Now, we can successfully separate the imperfection from $\tilde{\chi}$($\chi_{cphase}$.$\chi_{input}$). By compensating the input state imperfection, we were able to successfully characterize our time-bin qubit C-Phase gate itself. Figure~\ref{qptmatrixre}(c) shows the process matrix $\chi_{cphase}$, and the process fidelity $F_{cphase}$ is 97.1$\%$.

The average process fidelity, which is defined as the fidelity between the expected and actual output states averaged overall pure inputs, can be obtained with the formula $F_{avg}=(4F_{cphase}+1)/5$ \cite{gate_Bell,qptcnot,avgprocessfidelity}. With our experimental data, we obtained an average process fidelity of 97.7$\%$. This clearly demonstrated the switch was capable of a quantum gate operation. We also estimated the entangling capability ($\mathcal{C}\geq2F_{cphase}-1$) of our time-bin qubit C-Phase gate to be 0.94 \cite{hofmannprl2005}. It is important to note here that we did not subtract any noise counts when reconstructing the density and process matrices and calculating the fidelity in all our experiments, as is essential in on-line communications systems. TABLE~\ref{tab:table} summarizes the experimental results for the C-Phase gate.

\begin{figure*}[t]
\centering
\includegraphics[width=18cm]{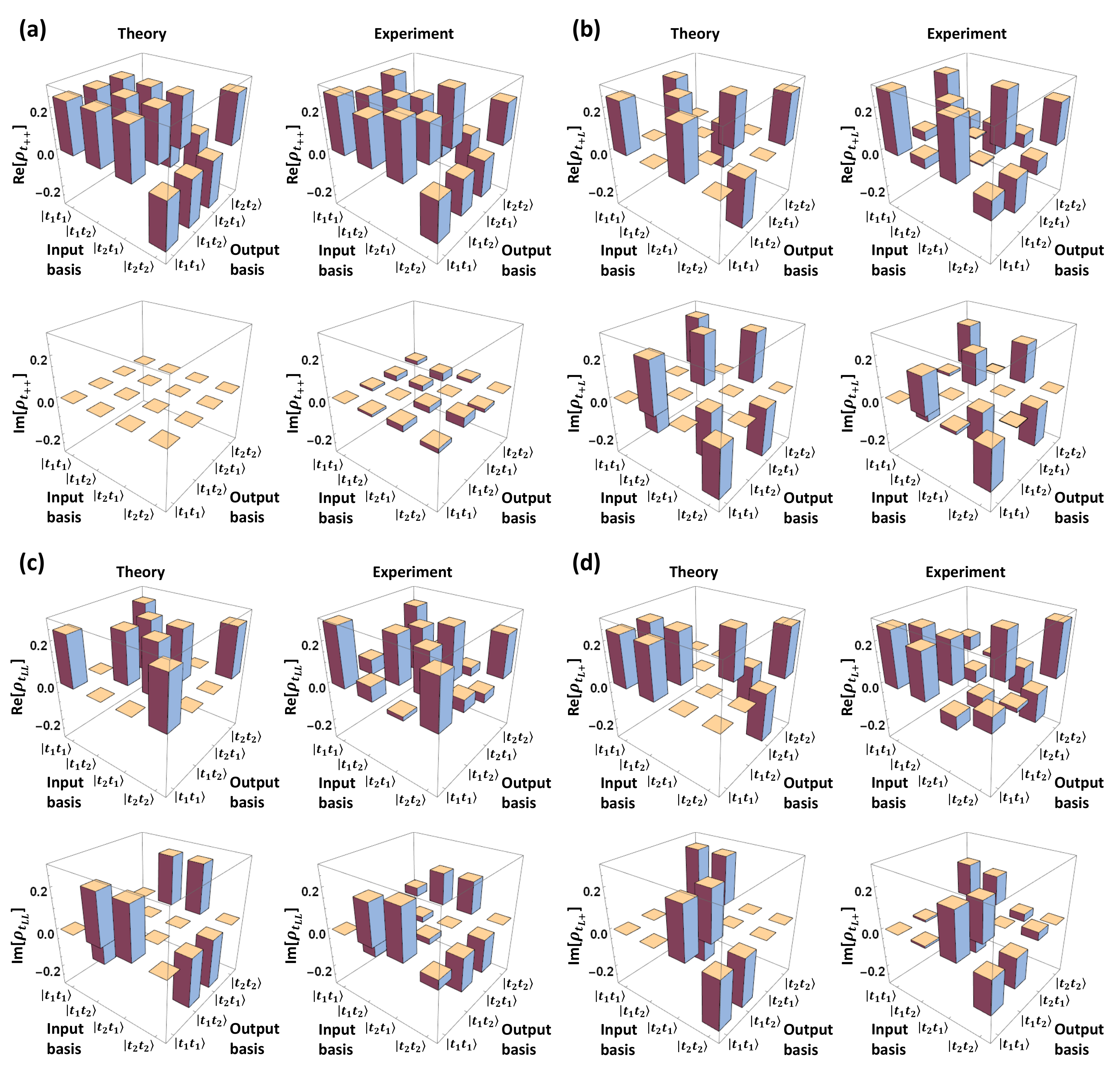}
\caption{Density matrices for the C-Phase gate entangling operation of inputs (a) $\left|++ \right\rangle$, (b) $\left|+L \right\rangle$, (c) $\left|LL \right\rangle$, and (d) $\left|L+ \right\rangle$ states. In each result, we show both the theoretical expectation on the left and the experimental result on the right. The upper and lower parts are the real and imaginary parts of each experimental density matrices.}
\label{fig:entangle}
\end{figure*}

\section{Time-bin qubit entanglement generation by operating the C-Phase gate}
The C-Phase gate can be used to generate maximally entangled states with specific input states, which are $\left|++ \right\rangle$, $\left|+L \right\rangle$, $\left|LL \right\rangle$, and $\left|L+ \right\rangle$ in the present experiment. For example, supposing $\left|LL \right\rangle=\frac{1}{2}(\left|t_1t_1 \right\rangle+i\left|t_1t_2 \right\rangle+i\left|t_2t_1 \right\rangle-\left|t_2t_2 \right\rangle)$ as the input state for the C-Phase gate operation, the output state is given by
\begin{eqnarray}
\frac{1}{\sqrt{2}}(\left|t_1L \right\rangle+i\left|t_2R \right\rangle), 
\end{eqnarray}
where $\left|R \right\rangle=\frac{1}{\sqrt{2}} (\left|t_1 \right\rangle -i \left|t_2 \right\rangle)$. In our experiment, Alice and Bob each chose the input state preparation method 4 shown in Fig.~\ref{expqpt}(b). After the C-Phase gate operation has been performed, Charlie and David perform QST and maximum likelihood estimation (MLE) \cite{qst, qst_timebin} to obtain physical density matrices. The theoretical and experimental density matrices after the C-Phase gate operation for target states $\left|++ \right\rangle$, $\left|+L \right\rangle$, $\left|LL \right\rangle$, and $\left|L+ \right\rangle$ are shown in Fig.~\ref{fig:entangle} \cite{statefidelity}. They show high state fidelities that violate Bell inequalities \cite{bell}.

\begin{figure}[t]
\centering 
\includegraphics[width=8.6cm]{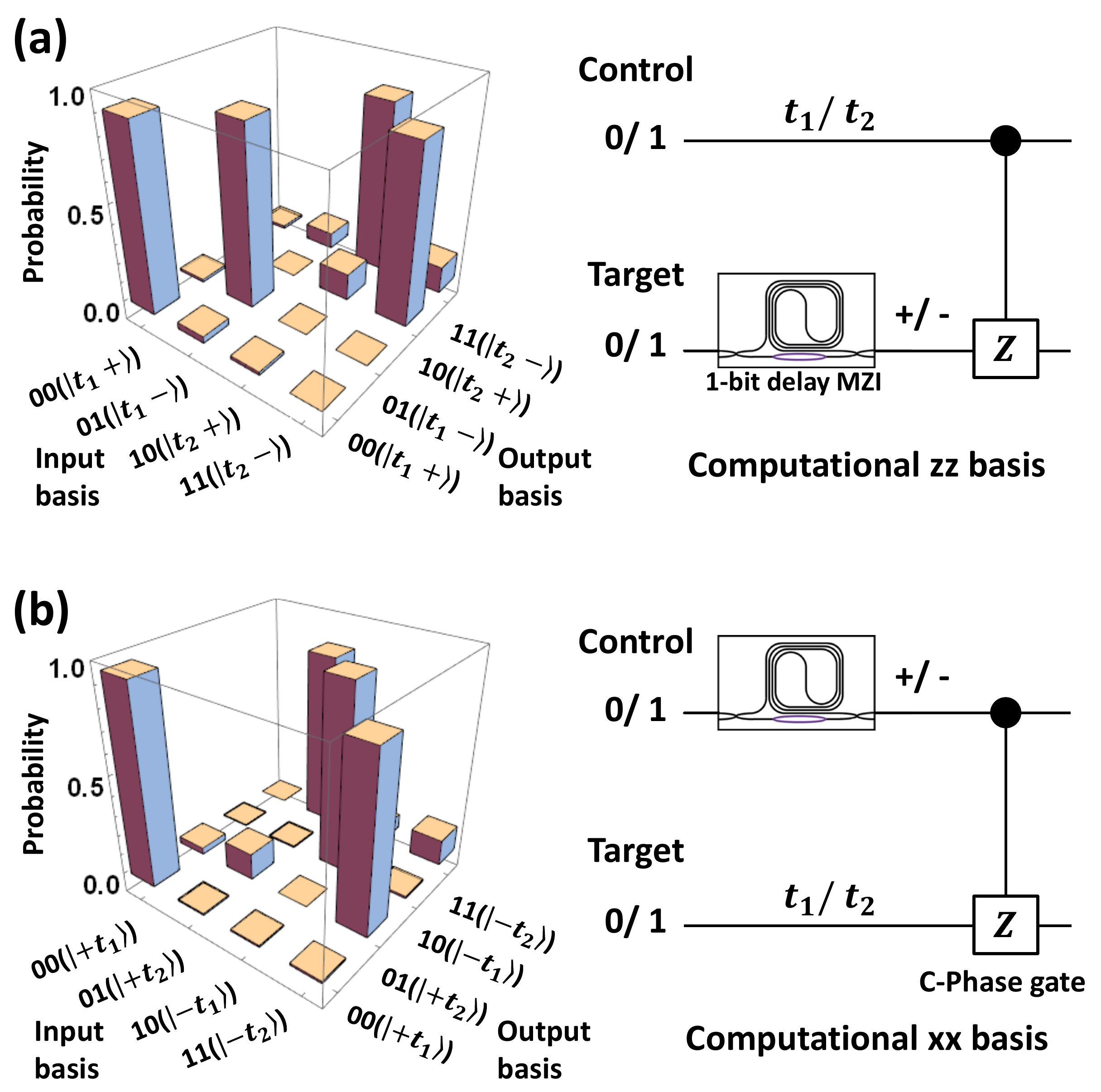}
\caption{Bar graph of our experimental CNOT gate for time-bin qubits in the (a) computational $zz$ and (b) $xx$ basis.}
\label{fig:cnot}
\end{figure}

\section{Implementation of a CNOT gate for time-bin qubits}
\label{sectioncnot}
The CNOT gate, another important two-qubit logic gate, flips the target qubit if and only if the control qubit is $\left|1 \right\rangle$ and it can be implemented by cascading Hadamard, C-Phase, and Hadamard gate operation \cite{Nielsen_Chuang}. In accordance with the scheme proposed in \cite{hofmannprl2005} and the demonstration given in \cite{cnot_okamoto}, we devised a CNOT gate for time-bin qubit operation with our C-Phase gate by employing complementary computational bases. First, Alice and Bob prepare the computational $zz$ basis of the gate $\left|0_z \right\rangle_c \equiv \left|t_1 \right\rangle_c$ and $\left|1_z \right\rangle_c \equiv \left|t_2 \right\rangle_c$ for the control qubit and $\left|0_z \right\rangle_t \equiv \left|+_z \right\rangle_t = \frac{1}{\sqrt{2}} (\left|t_1 \right\rangle + \left|t_2 \right\rangle)_t$ and $\left|1_z \right\rangle_t \equiv \left|-_z \right\rangle_t =\frac{1}{\sqrt{2}} (\left|t_1 \right\rangle - \left|t_2 \right\rangle)_t$ for the target qubit \cite{qst_timebin}, which is equivalent to performing the Hadamard gate operation on the target qubit. At the output ports, Charlie and David perform the coincidence measurement in the $zz$ basis and calculate the input-output probabilities of the CNOT gate operation, as shown in the Fig.~\ref{fig:cnot}(a). The logic fidelity $F_{zz}$, which is the average probability of obtaining the correct input-output probability, was 96.0 $\pm$ 7.2$\%$. We also measured the reversed CNOT gate operation using the complementary $xx$ basis of the gate $\left|0_x \right\rangle_c \equiv \left|+_x \right\rangle_c = \frac{1}{\sqrt{2}} (\left|t_1 \right\rangle + \left|t_2 \right\rangle)_c$ and $\left|1_x \right\rangle_c \equiv \left|-_x \right\rangle_c =\frac{1}{\sqrt{2}} (\left|t_1 \right\rangle - \left|t_2 \right\rangle)_c$ for the control qubit \cite{qst_timebin} and $\left|0_x \right\rangle_t \equiv \left|t_1 \right\rangle_t$ and $\left|1_x \right\rangle_t \equiv \left|t_2 \right\rangle_t$ for the target qubit, which means the target qubit acting on the control qubit. The experimental result is shown in Fig.~\ref{fig:cnot}(b); $F_{xx}$ was 97.8 $\pm$ 0.6$\%$. These two classical fidelities can be used to place bounds on the quantum process fidelity of the gate \cite{hofmannprl2005} as follows:
\begin{eqnarray}
F_{zz}+F_{xx}-1  \leq F_{CNOT} \leq \min\{F_{zz}, F_{xx}\}.
\label{cnot}
\end{eqnarray}
This means we do not need to perform full quantum process tomography to bound the efficiency of our CNOT gate. From our experimental results, we calculated the process fidelity $F_{CNOT}$ of the CNOT gate for time-bin qubits to be $0.94\leq F_{CNOT} \leq0.96$. This is as expected since our CNOT gate is composed of two imperfect Hadamard gates and an imperfect C-Phase gate. We can also estimate the minimal entangling capability of the logic gate by using the lower bound of the process fidelity \cite{hofmannprl2005}. Thus, the lower bound of the entangling capability is $\mathcal{C}\geq0.88$, if the minimal process fidelity of the gate is 0.94 \cite{cnot_okamoto}. These results confirm that like the C-Phase gate, our CNOT gate works well as an entangling gate. In TABLE~\ref{tab:table}, we summarize the experimental results for the CNOT gate.
\begin{table}
\caption{Summary of experimental results for time-bin qubit gates with uncertainties of $\sim$$\pm$0.06. The range for CNOT gate comes from the bounds indicated in Eq.~(\ref{cnot}).}
\label{tab:table}
\centering
\begin{tabular}{|c|c|c|}
\hline
Gate & Process fidelity & Entangling capability ($C$) \\
\hline 
C-Phase  & 0.971   & 0.94 \\
CNOT     & 0.94$\leq$$F_{CNOT}$$\leq$0.96   & 0.88 ($F_{CNOT}$=0.94) \\
\hline 
\end{tabular}
\end{table}   

\section{Discussion}
The results described above show that we have successfully realized high-fidelity C-Phase and CNOT gates. However, our experiment had several factors that may have decreased the state fidelity. The fluctuation of the splitting ratio of the 2$\times$2 switch, which arises from the DC bias drift of the lithium niobate waveguide modulator, may cause errors in the QST and fidelity of CNOT gate operation measurements. Moreover, because of the relatively large losses of the interferometers and optical switch, the QST required a long measurement time. In particular, QPT for two-qubit gates requires long measurement time for performing 16 QSTs, and this resulted in increased fluctuation of the C-Phase gate operation for time-bin qubits. Thanks to the PLC interferometers equipped with the additional MZI in the short arm, we were able to reduce the system loss and shorten the measurement time of the experiment. Otherwise, the input states are prepared by connecting different time-bin qubit preparation methods to input ports of the switch as shown in Fig.~\ref{expqpt}(b), which also causes system fluctuations. In addition, the accidental coincidence counts caused by dark counts and multiphoton emissions (estimated to be around $\sim$2$\%$ \cite{CAR}) from the correlated photon pair generation will also limit the state and gate process fidelities.

In our experiment, the total system loss is around -44 dB, which makes the experimental success probability is $\sim$0.04\% of the theory as shown in Sec.~\ref{concept}. If these technical issues can be resolved \cite{lowlosswaveguide}, we believe that quantum gates with much higher successful probability can be achieved. However, probabilistic logic gates based on postselection will have limited applications in small-scale quantum computation and communication. In particular, our gates will be useful for a quantum repeater based on small-scale error correction \cite{Bill_repeater}.

In \cite{qptcoherent}, Rahimi-Kashari et al. introduced the possibility of characterizing quantum logic gates with coherent state inputs. While the application of their scheme to two-qubit gates will constitute an important future work, we believe that characterizing our gates using the very states that we use in real applications is an essential task.

We would like to note that the C-Phase gate reported here together with single qubit operations are sufficient for creating any kind of quantum network based on time-bin qubits. By using active modulation, the Hadamard gate operation for time-bin qubits may also be demonstrated \cite{eoswitch}. The two-qubit gates could also be used to generate multipartite time-bin entangled states such as the GHZ state \cite{GHZNature} and cluster state \cite{eoswitch,cluster}. These technologies may open up new realms of application for time-bin qubits.

In conclusion, we demonstrated a C-Phase gate for time-bin qubits by using an active electro-optic 2$\times$2 optical switch. By adjusting the DC bias and RF modulation signal, the switch works as a TDBS operating as one-third beam splitter of the $t_2$ mode and passing the $t_1$ mode. Then, we reconstructed the positive process $\tilde{\chi}$ matrix by using QPT with 16 distinct inputs and performing MLE. Therefore, we could fully characterize our C-Phase gate for time-bin qubits and the process fidelity was 97.1$\%$. We also generated the time-bin entangled states by launching specific independent input qubits into the C-Phase gate operation. Moreover, we realized a time-bin qubit CNOT gate with a process fidelity greater than 94$\%$.


\begin{thebibliography}{1}
\bibitem{Nielsen_Chuang} M. A. Nielsen and I. L. Chuang, Quantum Computation and Quantum Information (Cambridge: Cambridge University Press, 2000).

\bibitem{chemistry} B. P. Lanyon, J. D. Whitfield, G. G. Gillett, M. E. Goggin, M. P. Almeida, I. Kassal, J. D. Biamonte, M. Mohseni, B. J. Powell, M. Barbieri, A. Aspuru-Guzik, and A. G. White, Towards quantum chemistry on a quantum computer, Nature Chemistry 2, 106 (2010). 

\bibitem{quantumcomputer} S. Debnath, M. M. Linke, C. Figgatt, K. A. Landsman, K. Wright, and C. Monroe, Demonstration of a small programmable quantum computer with atomic qubits, Nature 536, 63 (2016). 

\bibitem{steane_iontrap} A. Steane, The ion trap quantum information processor, Appl. Phys. B 64, 623 (1997). 

\bibitem{QST_superconducting_IBM} M. Steffen, M. Ansmann, R. C. Bialczak, N. Katz, E. Lucero, R. McDermott, M. Neeley, E. M. Weig, A. N. Cleland, and J. M. Martinis, Measurement of the Entanglement of Two Superconducting Qubits via State Tomography, Science 313, 1423 (2006). 

\bibitem{Li_Qdot_gate} X. Li, Y. Wu, D. Steel, D. Gammon, T. H. Stievater, D. S. Katzer, D. Park, C. Piermarocchi, and L. J. Sham, An All-Optical Quantum Gate in a Semiconductor Quantum Dot, Science 301, 809 (2003). 

\bibitem{Hanson_dot_RMP_2007} R. Hanson, L. P. Kouwenhoven, J. R. Petta, S. Tarucha, and L. M. K. Vandersypen, Spins in few-electron quantum dots, Rev. Mod. Phys. 79, 1217 (2007).

\bibitem{Zwanenburg_dot_RMP_2007} F. A. Zwanenburg, A. S. Dzurak, A. Morello, M. Y. Simmons, L. C. L. Hollenberg, G. Klimeck, S. Rogge, S. N. Coppersmith, and M. A. Eriksson, Silicon quantum electronics, Rev. Mod. Phys. 85, 961 (2013).

\bibitem{polspdc} P. G. Kwiat, K. Mattle, H. Weinfurter, A. Zeilinger, A. V. Sergienko, and Y. Shih, New High-Intensity Source of Polarization-Entangled Photon, Phys. Rev. Lett. 75, 4337 (1995). 

\bibitem{beamlikeLo} H. P. Lo, A. Yabushita, C. W. Luo, P. Chen, and T. Kobayashi, Beamlike photon-pair generation for two-photon interference and polarization entanglement, Phys. Rev. A 83, 022313 (2011).   

\bibitem{beamlike2x2} H. P. Lo, A. Yabushita, C. W. Luo, P. Chen, and T. Kobayashi, Beamlike photon pairs entangled by a 2$\times$2 fiber, Phys. Rev. A 84, 022301 (2011).   

\bibitem{qst_timebin} H. Takesue and Y. Noguchi, Implementation of quantum state tomography for time-bin entangled photon pairs, Opt. Express 17, 10976 (2009).

\bibitem{frequencybinqubit} J. M. Lukens and P. Lougovski, Frequency-encoded photonic qubits for scalable quantum information processing, Optica 4, 8 (2017).

\bibitem{gate_ion_trap} M. Riebe, K. Kim, P. Schindler, T. Monz, P. O. Schmidt, T. K. K\"{o}rber, W. H\"{a}nsel, H. H\"{a}ffner, C. F. Roos, and R. Blatt, Process Tomography of Ion Trap Quantum Gates, Phys. Rev. Lett. 97, 220407 (2006).   

\bibitem{cnot_gate_superconducting} J. H. Plantenberg, P. C. de Groot, C. J. P. M. Harmans, and J. E. Mooij, Demonstration of controlled-NOT quantum gates on a pair of superconducting quantum bits, Nature 447, 836 (2007).   

\bibitem{gate_superconducting} L. DiCarlo, J. M. Chow, J. M. Gambetta, L. S. Bishop, B. R. Johnson, D. I. Schuster, J. Majer, A. Blais, L. Frunzio, S. M. Girvin, and R. J. Schoelkopf, Demonstration of two-qubit algorithms with a superconducting quantum processor, Nature 460, 240 (2009).   

\bibitem{ion} D. Leibfried, B. DeMarco, V. Meyer, D. Lucas, M. Barrett, J. Britton, W. M. Itano, B. Jelenkovi$\acute{c}$, C. Langer, T. Rosenband, and D. J. Wineland, Experimental demonstration of a robust, high-fidelity geometric two ion-qubit phase gate, Nature 422, 412 (2003). 

\bibitem{phasequbit} T. Yamamoto, M. Neeley, E. Lucero, R. C. Bialczak, J. Kelly, M. Lenander, Matteo Mariantoni, A. D. O'Connell, D. Sank, H. Wang, M. Weides, J. Wenner, Y. Yin, A. N. Cleland, and John M. Martinis, Quantum process tomography of two-qubit controlled-Z and controlled-NOT gates using superconducting phase qubits, Phys. Rev. B 82, 184515 (2010).   

\bibitem{superconducting} H. J. Plantenberg, P. C. de Groot, C. J. P. M. Harmans, and J. E. Mooij, Demonstration of controlled-NOT quantum gates on a pair of superconducting quantum bits, Nature 447, 836 (2007).     

\bibitem{solid} Yu. A. Pashkin, T. Yamamoto, O. Astafiev, Y. Nakamura, D. V. Averin, and J. S. Tsai, Quantum oscillations in two coupled charge qubits, Nature 421, 823 (2003).

\bibitem{frequencybinCNOT} H.-H. Lu, J. M. Lukens, B. P. Williams, P. Imany, N. A. Peters, A. M. Weiner and P. Lougovski, A controlled-NOT gate for frequency-bin qubits, npj Quantum Inf. 5, 24 (2019).

\bibitem{KLM} E. Knill, R. Laflamme, G. J. Milburn, A scheme for efficient quantum computation with linear optics, Nature 409, 46 (2001). 
\bibitem{postselection} T. C. Ralph, A. G. White, W. J. Munro, and G. J. Milburn, Simple scheme for efficient linear optics quantum gates, Phys. Rev. A 65, 012314 (2001).   

\bibitem{cphasetheory} H. F. Hofmann and S. Takeuchi, Quantum phase gate for photonic qubits using only beam splitters and postselection, Phys. Rev. A 66, 024308 (2002).    

\bibitem{cnottheory} T. C. Ralph, N. K. Langford, T. B. Bell, and A. G. White, Linear optical controlled-NOT gate in the coincidence basis, Phys. Rev. A 65, 062324 (2002).   

\bibitem{gate_photon_OBrien} J. L. O'Brien, G. J. Pryde, A. G. White, T. C. Ralph, and D. Branning, Demonstration of an all-optical quantum controlled-NOT gate, Nature 426, 264 (2003). 

\bibitem{cphase_pdbs} N. Kiesel, C. Schmid, U. Weber, R. Ursin, and H. Weinfurter, Linear Optics Controlled-Phase Gate Made Simple, Phys. Rev. Lett. 95, 210505 (2005).   

\bibitem{gate_Bell} N. K. Langford, T. J. Weinhold, R. Prevedel, A. Gilchrist, J. L. O'Brien, G. J. Pryde, and A. G. White, Demonstration of a Simple Entangling Optical Gate and Its Use in Bell-State Analysis, Phys. Rev. Lett. 95, 210504 (2005).  

\bibitem{kok_RMP} P. Kok, W. J. Munro, K. Nemoto, T. C. Ralph, J. P. Dowling, and G. J. Milburn, Linear optical quantum computing with photonic qubits, Rev. Mod. Phys. 79, 135 (2007).  

\bibitem{qinternet} H. J. Kimble, The quantum internet, Nature 453, 1023 (2008).  

\bibitem{Bill_repeater} W. J. Munro, A. M. Stephens, S. J. Devitt, K. A. Harrison, K. Nemoto, Quantum communication without the necessity of quantum memories, Nat. Photon. 6, 777 (2012).

\bibitem{time-energy_Brendel_1999} J. Brendel, N. Gisin, W. Tittel, and H. Zbinden, Pulsed Energy-Time Entangled Twin-Photon Source for Quantum Communication, Phys. Rev. Lett. 82, 2594 (1999).   
\bibitem{time-energy_Gisin_2007} N. Gisin and R. Thew, Quantum communication, Nat. Photon. 1, 165 (2007). 

\bibitem{Honjo_timebin_QKD} T. Honjo, S. W. Nam, H. Takesue, Q. Zhang, H. Kamada, Y. Nishida, O. Tadanaga, M. Asobe, B. Baek, R. Hadfield, S. Miki, M. Fujiwara, M. Sasaki, Z. Wang, K. Inoue, Y. Yamamoto, Long-distance entanglement-based quantum key distribution over optical fiber, Opt. Express 16, 19118 (2008). 

\bibitem{cphasetimebin} H. P. Lo, T. Ikuta, N. Matsuda, T. Honjo, and H. Takesue, Entanglement generation using a controlled-phase gate for time-bin qubits, Appl. Phys. Exp. 11 092801 (2018).   

\bibitem{takesue_switch} H. Takesue, Entangling time-bin qubits with a switch, Phys. Rev. A 89, 062328 (2014). 


\bibitem{CAR} H. Takesue and K. Shimizu, Effects of multiple pairs on visibility measurements of entangled photons generated by spontaneous parametric processes, Opt. Commun. 283, 276 (2010). 

\bibitem{qst} D. F. V. James, P. G. Kwiat, W. J. Munro, and A. G. White, Measurement of qubits, Phys. Rev. A 64, 052312 (2001).   

\bibitem{takesue_teleport_2015} H. Takesue, S. D. Dyer, M. J. Stevens, V. Verma, R. P. Mirin, and S. W. Nam, Quantum teleportation over 100 km of fiber using highly efficient superconducting nanowire single-photon detectors, Optica 2, 832 (2015). 

\bibitem{takesue_plc} H. Takesue and K. Inoue, Generation of 1.5--$\mu$m band time-bin entanglement using spontaneous fiber four-wave mixing and planar light-wave circuit interferometers, Phys. Rev. A 72, 041804(R) (2005). 





\bibitem{QPT_white2007} A. G. White, A. Gilchrist, G. J. Pryde, J. L. O'Brien, M. J. Bremner, and N. K. Langford, Measuring two-qubit gates, J. Opt. Soc. Am. B 24, 172 (2007). 

\bibitem{qptcnot}	J. L. O'Brien, G. J. Pryde, A. Gilchrist, D. F. V. James, N. K. Langford, T. C. Ralph, and A. G. White, Quantum Process Tomography of a Controlled-NOT gate, Phys. Rev. Lett. 93, 080502 (2004).   

\bibitem{mle} M. Howard, J. Twamley, C. Wittmann, T. Gaebel, F. Jelezko, and J. Wrachtrup, Quantum process tomography and Linblad estimation of a solid-state qubit, New J. Phys. 8, 33 (2006).   

\bibitem{error} Errors of process fidelity is calculated from Poissonian counting statistics.



\bibitem{avgprocessfidelity} A. Gilchrist, N. K. Langford, and M. A. Nielsen, Distance measures to compare real and ideal quantum processes, Phys. Rev. A 71, 062310 (2005).   

\bibitem{hofmannprl2005} H. F. Hofmann, Complementary Classical Fidelities as an Efficient Criterion for the Evaluation of Experimentally Realized Quantum Operations, Phys. Rev. Lett. 94, 160504 (2005).  

\bibitem{statefidelity} In our experiment, the average state fidelity of 16 density matrices to the pure states was 93.6 $\pm$ 5.8$\%$. The average state fidelity of the four specific input states to the pure states was 90.1 $\pm$ 5.1$\%$. The state fidelities to the target entangled states for inputs $\left|++ \right\rangle$, $\left|+L \right\rangle$, $\left|LL \right\rangle$, and $\left|L+ \right\rangle$ were $F_{++}$= 91.4 $\pm$ 10.5$\%$, $F_{+L}$= 94.8 $\pm$ 2.7$\%$, $F_{LL}$= 85.1 $\pm$ 4.2$\%$, and $F_{L+}$= 88.9 $\pm$ 2.9$\%$, respectively.



\bibitem{bell} C. H. Bennett, G. Brassard, S. Popescu, B. Schumacher, J. A. Smolin, and W. K. Wootters, Purification of Noisy Entanglement and Faithful Teleportation via Noisy Channels, Phys. Rev. Lett. 76, 722 (1996).

\bibitem{cnot_okamoto} R. Okamoto, H. F. Hofmann, S. Takeuchi, and K. Sasaki, Demonstration of an Optical Quantum Controlled-NOT Gate without Path Interference, Phys. Rev. Lett. 95, 210506 (2005).   

\bibitem{lowlosswaveguide} V. Ramaswamy, R.C. Alferness, and M. Divino, High efficiency single-mode fibre to Ti:LiNbO3 waveguide coupling, Electron. Lett. 18,30–31 (1982).   

\bibitem{qptcoherent} S. Rahimi-Keshari, A. Scherer, A. Mann, A. T. Rezakhani, A. I. Lvovsky, and B. C. Sanders, “Quantum process tomography with coherent states,” New J. Phys. 13, 013006 (2011).

\bibitem{eoswitch} Y. Soudagar, F. Bussi\'{e}res, G. Berl\'{i}n, S. Lacroix, J. M. Fernandez, and N. Godbout, Cluster-state quantum computing in optical fibers, J. Opt. Soc. Am. B 24, 226 (2007).   


\bibitem{GHZNature} J.-W. Pan, D. Bouwmeester, M. Daniell, H. Weinfurter, and A. Zeilinger, Experimental test of quantum nonlocality in three-photon Greenberger–Horne–Zeilinger entanglement, Nature 403, 515 (2000).  

\bibitem{cluster} M. A. Nielsen, Optical Quantum Computation Using Cluster States, Phys. Rev. Lett. 93, 040503 (2004).   


\end{thebibliography}
\end{document}